\documentclass[10pt,a4paper]{article}
\usepackage[T2A]{fontenc}
\usepackage[utf8]{inputenc}
\usepackage[english]{babel}
\usepackage{amssymb,graphicx}
\usepackage{amsmath,amsfonts}
\usepackage{amsthm}
\usepackage{framed}
\usepackage{fullpage}
\usepackage{microtype}
\usepackage{hyperref}

 \newcommand{\Tr}{\mathop{\mathrm{Tr}}\nolimits}

\newcommand{\normord}[1]{\,\mathopen{:}\,#1\,\mathclose{:}\,}

\title{\Large{\textbf{$3d$ field theory, plane partitions and triple
      Macdonald polynomials}}} \date{}
\author{Yegor Zenkevich\thanks{yegor.zenkevich@gmail.com}\\
  {\small\textit{Dipartimento di Fisica, Universit\`a di Milano
      Bicocca, Piazza della Scienza 3, I-20126 Milano, Italy}}\\
  {\small\textit{INFN, sezione di Milano-Bicocca, I-20126 Milano, Italy}}\\
  {\small\textit{ITEP, Moscow 117218, Russia}}} \date{}

\begin{document}
\maketitle
\vspace{-50ex}
\begin{flushright}
  ITEP-TH-45/17
\end{flushright}
\vspace{43ex}

\begin{abstract}
  We argue that MacMahon representation of Ding-Iohara-Miki (DIM)
  algebra spanned by plane partitions is closely related to the
  Hilbert space of a $3d$ field theory. Using affine matrix model we
  propose a generalization of Bethe equations associated to DIM
  algebra with solutions also labelled by plane partitions. In a
  certain limit we identify the eigenstates of the Bethe system as new
  \emph{triple} Macdonald polynomials depending on an infinite number
  of families of time variables. We interpret these results as first
  hints of the existence of an integrable $3d$ field theory, in which
  DIM algebra plays the same role as affine algebras in $2d$ WZNW
  models.
\end{abstract}

\section{Introduction}
\label{sec:introduction}

Nekrasov functions~\cite{Nekrasov} obtained by applying the
localization technique to supersymmetric gauge theories in four, five
and six dimensions effectively sum up instanton contributions to their
partition functions. Instanton configurations fixed under toric action
of $U(1)^2_{t_1,t_2}$ are labelled by tuples of Young diagrams, which
technically arise as ideals in the ring of polynomials in two
variables, or as residues of the LMNS integral~\cite{LMNS}. Young
diagrams naturally remind one of the basis in the Hilbert space of a
$2d$ field theory. It turns out that this connection can indeed be
made precise leading to the famous AGT correspondence~\cite{AGT}. Each
term in the instanton sum over partitions corresponds to a particular
state in the Hilbert space (Verma module) of a $2d$ CFT. The states
featuring in the correspondence are eigenstates of a certain family of
CFT integrals of motion~\cite{Belavin} and are naturally labelled by
tuples of Young diagrams. The Hilbert space of the $2d$ theory is a
representation of $W_N$-algebra, and is identified with the
equivariant cohomology of the gauge theory instanton moduli space.

In this paper we develop the basics of a similar, though not yet as
precise, identification for \emph{plane} partitions, i.e.\ $3d$ Young
diagrams. To this end in sec.~\ref{sec:3d-free-scalar} we study $3d$
free scalar field theory and identify (chiral half of) its Hilbert
space as the space of plane partitions. We then consider in
sec.~\ref{sec:affine-matrix-models}, instead of the LMNS integral, the
affine matrix model, which poles are labelled by plane partitions. We
derive the quantum spectral curve for this model and, taking its
Nekrasov-Shatashvili (NS) limit~\cite{NS}, obtain certain Baxter TQ
equation. A particular case of this Baxter equation already appeared
in the literature~\cite{ILW, FJMM} as an equation determining the
eigenstates of the integrals of motion associated to the
Ding-Iohara-Miki algebra
$U_{t_1,t_2}(\widehat{\widehat{\mathfrak{gl}}}_1)$. In this restricted
case the solutions were labelled by $K$-tuples of Young diagrams, so
that the eigenstates spanned a tensor product of $K$ Fock
representations ($K$ $2d$ fields). However, the slight generalization
of the Baxter equation obtained from the affine matrix model has
solutions labelled by $K$-tuples of $3d$ Young diagrams. The states
corresponding to these solutions, therefore, span the tensor product
of what is called the \emph{MacMahon representations} of
$U_{t_1,t_2}(\widehat{\widehat{\mathfrak{gl}}}_1)$. It is natural to
associate this representation with the states of a $3d$ field.

Finally, in sec.~\ref{sec:triple-macd-polyn} we give a concrete
construction of a basis in the MacMahon representation of DIM as a
subrepresentation in an infinite tensor product of Fock
representations. The basis, which we call the family of \emph{triple}
Macdonald polynomials, is formed by the eigenstates of the Heisenberg
subalgebra of DIM algebra. This basis can also be obtained in a
certain limit of DIM Baxter equation mentioned above. It is tempting
to speculate that this construction implies the existence of an
\emph{integrable} $3d$ field theory, though we are currently very far
from giving its concrete description. Ideologically, this connection
might be justified by noticing that loop algebras (affine, Virasoro or
$W_N$) naturally appear in $2d$ CFT, while DIM algebra is essentially
a \emph{double} loop algebra, and thus should be associated to a $3d$
theory.

We use the standard notation for Young diagrams $Y = [Y_1, Y_2,\ldots,
Y_m]$ with $Y_1 \geq Y_2 \geq \cdots \geq Y_m$ throughout the
paper. Plane partition $\pi = [\pi^{(1)}, \pi^{(2)}, \ldots,
\pi^{(n)}]$ is built from the ``non-increasing'' sequence of Young
diagrams $\pi^{(1)} \supset \pi^{(2)} \supset \cdots \supset \pi^{(n)} $.

\section{$3d$ massless free scalar and $3d$ partitions}
\label{sec:3d-free-scalar}

In this section we demonstrate how the Hilbert space of the simplest
possible $3d$ field theory, the massless free scalar, is related to
the space of plane partitions. We start by recalling the prototypical
example of $2d$ free scalar, which Hilbert space factorizes into two
chiral parts, each of which is a Fock space $\mathcal{F}$ with basis
labelled by ordinary partitions (Young diagrams). We then analyze the
$3d$ setup using similar arguments. In particular, we introduce an
analogue of the holomorphic factorization and observe that the Hilbert
space of the free scalar in $3d$ (after certain reduction) factorizes
into a product of two \emph{MacMahon} spaces spanned by \emph{plane}
partitions.

Our considerations are very elementary and the free scalar is only a
toy model for more interesting $3d$ field theories (see e.g.\ the
recent work~\cite{Aganagic:2017tvx}). Still, we believe that the
general structure of the Hilbert space will remain the same as in the
free case, especially if one considers \emph{integrable} $3d$ models.

\subsection{Warm-up: $2d$ free scalar}
\label{sec:warm-up:-2d}
One way to quantize a classical model is to quantize the space of
solutions to its classical equations of motion. One can also interpret
it as the quantization of the space of initial data, since each
solution is uniquely determined by the initial conditions\footnote{We
  assume that the Cauchy problem for the classical model is well
  posed.}. In the mechanical systems the quantization amounts to
replacing the coordinates on the space of solutions (e.g.\ initial
coordinates and momenta) with non-commuting operators. In the field
theory this procedure is the same in principle, though the space under
consideration is infinite-dimensional, which may require more subtle
treatment. However, for the free fields, no technical problems arise
and the quantization can be carried out straightforwardly.

We will work in the Euclidean signature and the \emph{radial}
coordinate will play the role of time. The classical equations of
motion for a $2d$ massless free scalar read
\begin{equation}
  \label{eq:12}
  (\partial_x^2 + \partial_y^2) \phi(x,y) = 0
\end{equation}
Changing to the radial coordinates $(r,\varphi)$ we get
\begin{equation}
  \label{eq:13}
  \left[(r \partial_r)^2 + \partial_{\varphi}^2 \right]
  \phi(r,\varphi) = 0
\end{equation}
One can find a basis of solutions by separating the variables, i.e.\
$\phi(r,\varphi) = f(r) g(\varphi)$. From the fact that $g(\varphi)$
needs to be single-valued we get
\begin{equation}
  \label{eq:14}
  g(\varphi) = e^{i m \varphi} ,\qquad m \in \mathbb{Z}.
\end{equation}
The equation for $f(r)$ is then
\begin{equation}
  \label{eq:15}
  (r\partial_r)^2 f(r) = m^2 f(r),
\end{equation}
and thus
\begin{equation}
  \label{eq:18}
  f(r) = r^{\pm m}.
\end{equation}
At this point we need to specify on which part of the $2d$ plane the
theory lives. We assume it to be a large disk $D_2$ around $r =
0$. Therefore, we can only consider \emph{positive} powers of $r$, and
thus $f(r) = r^{|m|}$. We ignore the \emph{zero mode} solutions $f(r)
= 1$ and $f(r) = \ln r$ corresponding to $m =0$. One can assume that
they are killed by the boundary conditions on $\partial D_2$. We will
observe a similar situation when we come to the $3d$ case: part of the
solutions will need to be discarded. One can also notice that the
space of solutions which we have thrown away is exactly the space of
solutions to the free scalar equations of motion \emph{in one
  dimension less.}  Indeed the functions $1$ and $\ln r$ correspond to
initial position and momentum of a \emph{particle on a line.} In
string theory this coordinate and momentum correspond to the motion of
the string center of mass, while other solutions represent the
harmonics.

Regular solutions to the free field equations of motion are thus
labelled by $m \in \mathbb{Z}\backslash \{0\}$:
\begin{equation}
  \label{eq:16}
  \phi_m(r,\varphi) = r^{|m|} e^{i m \varphi}.
\end{equation}

To quantize the space of solutions we need to introduce the
coordinates. Those are the coefficients $a_{-n}$ and $\bar{a}_{-n}$ in
the expansion of a general field $\phi(r,\varphi)$ in the
basis~\eqref{eq:16}:
\begin{equation}
  \label{eq:17}
  \phi(r,\varphi) = \sum_{m \geq 1} \frac{1}{m} a_{-m} r^m
  e^{im \varphi}  + \sum_{m \geq 1} \frac{1}{m} \bar{a}_{-m} r^m e^{-im \varphi} 
\end{equation}
They constitute in fact only half of the coordinates --- the so-called
positive-frequency parts. The other half corresponds to the solutions
which are singular at $r=0$, but are instead regular at $r=\infty$.
If we introduce the \emph{complex structure} on the $2d$ plane, so
that the complex coordinates read $z = r e^{i\varphi}$, $\bar{z}= r
e^{-i \varphi}$, then the coordinates $a_{-n}$ and $\bar{a}_{-n}$
correspond to \emph{holomorphic} and \emph{antiholomorphic} fields
respectively.

The operators corresponding to the coordinates $a_{-n}$ and
$\bar{a}_{-n}$ commute between themselves and are the \emph{creation}
operators. They can be used to build all states of the model out of
the vacuum state $|\varnothing \rangle$. After the choice of complex
structure is made, the Hilbert space $\mathcal{H}_{\mathrm{tot}}^{2d}$
of the model, therefore, splits into a tensor product
$\mathcal{H}^{2d} \otimes \bar{\mathcal{H}}^{2d}$ of holomorphic and
antiholomorphic states:
\begin{gather}
  \label{eq:19}
  \mathcal{H}^{2d} = \bigoplus_{m_1 \geq m_2 \geq \cdots \geq m_k}
  \mathbb{C} a_{-m_1} a_{-m_2} \cdots a_{-m_k} |\varnothing\rangle\\
  \bar{\mathcal{H}}^{2d} = \bigoplus_{m_1 \geq m_2 \geq \cdots \geq m_k}
  \mathbb{C} \bar{a}_{-m_1} \bar{a}_{-m_2} \cdots \bar{a}_{-m_k}
  |\varnothing\rangle
\end{gather}
We can introduce the \emph{grading} $L_0$ on $\mathcal{H}$ by
assigning degree $m$ to $a_{-m}$. This choice of the degree
corresponds to the action of the dilatation operator $z \partial_z$ on
the solutions. Similar grading $\bar{L}_0 \simeq
\bar{z} \partial_{\bar{z}}$ acts on $\bar{\mathcal{H}}^{2d}$.

The states of $\mathcal{H}^{2d}$ are labelled by Young diagrams
$\{m_1, m_2,\ldots, m_k\}$ (notice that the sequence is
non-increasing). The partition function of the model can be written as
a product of two Dedekind eta-functions:
\begin{equation}
  \label{eq:20}
  Z_{2d}'(q,\bar{q}) = \Tr_{\mathcal{H}_{\mathrm{tot}}^{2d}} \left( q^{L_0}
    \bar{q}^{\bar{L}_0}\right) = \Tr_{\mathcal{H}^{2d}} \left( q^{L_0}
  \right) \Tr_{\bar{\mathcal{H}}^{2d}} \left( \bar{q}^{\bar{L}_0}
  \right) = \prod_{m\geq 1} \frac{1}{1-q^m} \frac{1}{1-\bar{q}^m} =
  \sum_{Y \in 2d \text{ YD}} q^{|Y|}\sum_{W\in 2d \text{ YD}} \bar{q}^{|W|}.
\end{equation}
where $Y$ and $W$ are $2d$ Young diagrams and $|Y|$ denotes the total
number of boxes in $Y$. The coordinates of the zero modes, that we
have thrown away earlier (hence a prime in $Z_{2d}'$) would have
contributed to the partition function by an overall factor of the form
$q^P$, which can be thought of as the partition function of a particle
with a given momentum, or as the contribution of the center of mass
coordinate.

Let us recapitulate the result in the $2d$ case. The solutions of the
equations of motion can be split into two parts: those regular at
$r=0$ and those regular at $r=\infty$. The first subspace corresponds
to the creation operators. There is also a small subspace of zero
modes, which needs to be discarded (or accounted for
separately). After introduction of the complex structure, creation
operators further split into holomorphic and antiholomorphic subsets,
each of which produces the Hilbert space with basis labelled by Young
diagrams. The partition function is the product of two generating
functions for the number of Young diagrams.

\subsection{$3d$ free scalar}
\label{sec:3d-free-scalar-1}
We will now consider the $3d$ free scalar and analyze its quantization
in the same spirit as the $2d$ case in the previous section. The
equation of motion reads
\begin{equation}
  \label{eq:21}
    (\partial_x^2 + \partial_y^2 + \partial_t^2) \phi(x,y,t) = 0
\end{equation}
or in the spherical coordinates:
\begin{equation}
  \label{eq:22}
  \left[(r \partial_r)^2 + r \partial_r + \Delta_{S^2} \right]
  \phi(r,\theta,\varphi) = 0
\end{equation}
where $\Delta_{S^2}$ is the Laplacian operator on a unit
sphere. Separating the variables we get $\phi(r,\theta,\varphi) = f(r)
g(\theta, \varphi)$, where $g(\theta, \varphi)$ should be an
eigenfunction of $\Delta_{S^2}$. It is thus given by the spherical
harmonics:
\begin{gather}
  \label{eq:23}
  g(\theta, \varphi) = Y_{j,m}(\theta, \varphi),\qquad j \in
  \mathbb{Z}_{\geq 0}, \quad -j \leq m \leq j,\\
  \Delta_{S^2} Y_{j,m}(\theta, \varphi) = - j(j + 1) Y_{j,m}(\theta,
  \varphi).
\end{gather}
The corresponding function $f(r)$ satisfies
\begin{equation}
  \label{eq:24}
  r\partial_r (r\partial_r + 1) f(r) = j(j+1) f(r),
\end{equation}
and thus
\begin{equation}
  \label{eq:25}
  f(r) = r^j \quad \text{or}\quad r^{-1-j}
\end{equation}
Only the first choice is regular at $r=0$, so the
solutions are given by
\begin{equation}
  \label{eq:26}
  \phi_{j,m}(r,\varphi) = r^j Y_{j,m}(\theta, \varphi),\qquad j \in
  \mathbb{Z}_{\geq 0}, \quad -j \leq m \leq j.
\end{equation}
Similarly to the $2d$ case we would like to discard certain solutions
and split the remaining ones using some analogue of the complex
structure\footnote{The relevant mathematical structure here is
  probably the \emph{contact} structure represented by the standard
  one-form $\lambda = dt + xdy- ydx = dt - \frac{i}{2} (\bar{z}dz - z
  d \bar{z}) $ in $\mathbb{R}^3$. Notice that $d\lambda = 2dx \wedge
  dy = i dz \wedge d\bar{z}$ represents the complex (or symplectic)
  structure on the constant $t$ slices of $\mathbb{R}^3$.}. The
natural choice of the solutions to be discarded are those with
$m=0$. One way to justify this choice is to require the solutions to
vanish at the $t$ axis. This condition can be enforced on the distant
boundary of the ball $B_3$. There is only one discarded solution for
any given $j$, so they can be thought of as corresponding to a $2d$
chiral scalar. Thus, the discarded solutions follow the pattern
established in $2d$: they correspond to the states of the field in one
dimension \emph{less} than the original model.

We split the remaining solutions into those with positive and negative
$m$ calling them, by gross abuse of the terminology, holomorphic and
antiholomorphic respectively. The coordinates on the space of
solutions are the coefficients in the expansion:
\begin{equation}
  \label{eq:27}
  \phi(r,\theta,\varphi) = \sum_{j \geq} \sum_{m=1}^j a_{-(j, m)} r^j
  Y_{j,m}(\theta, \varphi) + \sum_{j \geq} \sum_{m=1}^j \bar{a}_{-(j, m)} r^j
  Y_{j,-m}(\theta, \varphi)
\end{equation}
The Hilbert space $\mathcal{H}_{\mathrm{tot}}^{3d}$ is again a tensor
product of two factors $\mathcal{H}^{3d} \otimes
\bar{\mathcal{H}}^{3d}$. The space of holomorphic states
$\mathcal{H}^{3d}$ is spanned by the vectors created from the vacuum
$|\varnothing \rangle$ by the strings of operators $a_{-(j,m)}$:
\begin{gather}
  \label{eq:29}
  \mathcal{H}^{3d} = \bigoplus_{
    \begin{smallmatrix}
      j_1 \geq j_2 \geq \cdots \geq j_k\\
      m_1 \geq m_2 \geq \cdots \geq m_k
    \end{smallmatrix}
}
  \bigoplus_{m_1=1}^{j_1} \cdots \bigoplus_{m_k=1}^{j_k}
  \mathbb{C} a_{-(j_1,m_1)} a_{-(j_2,m_2)} \cdots a_{-(j_k,m_k)} |\varnothing\rangle\\
  \bar{\mathcal{H}}^{3d} = \bigoplus_{\begin{smallmatrix}
      j_1 \geq j_2 \geq \cdots \geq j_k\\
      m_1 \geq m_2 \geq \cdots \geq m_k
    \end{smallmatrix}} \bigoplus_{m_1=1}^{j_1} \cdots \bigoplus_{m_k=1}^{j_k}
  \mathbb{C} \bar{a}_{-(j_1,m_1)} \bar{a}_{-(j_2,m_2)} \cdots
  \bar{a}_{-(j_k,m_k)} |\varnothing\rangle
\end{gather}
The grading on $\mathcal{H}^{3d}$ corresponds to the the total power
of $r$ in the solution, so that $a_{-(j,m)}$ has degree $j$.

Quite remarkably, the states on a given level in $\mathcal{H}^{3d}$
are in one to one correspondence with \emph{plane partitions:}
\begin{equation}
  \label{eq:30}
  \begin{array}{c|c|c}
    & \text{Plane partitions} & \text{States in $\mathcal{H}$} \\
    \hline
    \text{deg} = 0 & \varnothing & |\varnothing \rangle \\
    \hline
    \text{deg} = 1 & [[1]] & a_{-(1,1)}|\varnothing \rangle\\
    \hline
    \text{deg} = 2 &
    [[1,1]],\quad [[2]],\quad [[1],[1]]
    &
    a^2_{-(1,1)}|\varnothing \rangle,\quad a_{-(2,1)}|\varnothing \rangle,\quad
    a_{-(2,2)}|\varnothing \rangle\\
    \hline
    \text{deg} = 3 & \begin{array}{c}
      [[1,1,1]],\quad [[2,1]],\quad [[3]]\\
      {} [[1,1],[1]],\quad [[2],[1]],\quad [[1],[1],[1]]
    \end{array}  &\begin{array}{c}
    a_{-(1,1)}^3|\varnothing \rangle,\quad a_{-(2,1)}
    a_{-(1,1)}|\varnothing \rangle, \quad a_{-(3,1)}|\varnothing
    \rangle \\
    a_{-(2,2)}a_{-(1,1)}|\varnothing \rangle,\quad a_{-(3,2)}
    |\varnothing \rangle, \quad a_{-(3,3)}|\varnothing \rangle
  \end{array}
\end{array}
\end{equation}
The pattern is clear, though explicit combinatorics of the mapping
partitions to polynomials of $a_{-(j,m)}$ will not concern us here.

The $3d$ partition function (with $m=0$ states removed) is thus equal
to the product of two \emph{MacMahon} functions\footnote{Hints of the
  connection between $3d$ scalar and MacMahon representation were
  mentioned in~\cite{Awata:1994tf}.}:
\begin{equation}
  \label{eq:28}
  Z_{3d}'(q,\bar{q}) = \Tr_{\mathcal{H}_{\mathrm{tot}}^{3d}} \left( q^{L_0}
    \bar{q}^{\bar{L}_0} \right) =  \Tr_{\mathcal{H}^{3d}} \left( q^{L_0}
  \right) \Tr_{\bar{\mathcal{H}}^{3d}} \left( \bar{q}^{\bar{L}_0}
  \right) = \prod_{j \geq 1} \frac{1}{(1-q^j)^j}
  \frac{1}{(1-\bar{q}^j)^j} = \sum_{\pi\in 3d \text{ YD}} q^{|\pi|}
  \sum_{\xi \in 3d \text{ YD}} \bar{q}^{|\xi|}
\end{equation}
Adding back the solutions, which we have discraded corresponds to
multiplying the partition function by an additional eta-function.

What we see is that a particular splitting of the states of the free
boson in $3d$ gives the analogue of holomorphic factorization for the
$2d$ boson and that the states of the Hilbert space are labelled by
(pairs of) plane partitions $\pi$.

\subsection{MacMahon representation $M_c(u)$ of $W_{1+\infty}$ and DIM
  algebras}
\label{sec:macm-repr-w_1+}
The description of the chiral Hilbert space $\mathcal{H}^{3d}$ spanned
by plane partitions $\pi$ in terms of creation operators $a_{-(j,m)}$
might look a little unconventional. In this section we recall a very
similar description of the MacMahon representation of the
$W_{1+\infty}$ algebra discovered in~\cite{Kac:1993zg} (see also the
review~\cite{ Awata:1994tf}). In this way we make a connection between
the $W_{1+\infty}$ algebra (and as we will see, DIM algebra in
general) and the states of the $3d$ free scalar.

DIM algebra can be thought of as a $t$-deformation of the Lie algebra
$W_{1+\infty}$. This Lie algebra is the central extension\footnote{Our
  choice of the central extension is equivalent to the more
  conventional choice leading to the usual central extension in the
  Virasoro subalgebra.} of the algebra of difference operators
$W_{(n,m)} \simeq q^{-\frac{nm}{2}} z^n q^{m z \partial_z}$:
\begin{equation}
  \label{eq:31}
  W_{1+\infty}:\qquad [W_{n,m}, W_{k,l}] = \left( q^{\frac{mk-nl}{2}}-
    q^{-\frac{mk-nl}{2}}\right) W_{n+k,m+l} + (nc_1 + m c_2)
  \delta_{n+k,0} \delta_{m+l,0}, \qquad n,m,k,l \in \mathbb{Z},
\end{equation}
where $c_1$ and $c_2$ are two central charges. One can equivalently
understand $W_{1+\infty}$ as the algebra of maps from the quantum
torus $(z, q^{z\partial_z})$ to $\mathbb{C}$. Notice that the
relations of the algebra are manifestly invariant under the action of
the $SL(2,\mathbb{Z})$ automorphism group, which is the mapping class
group of the (quantum) torus. In particular, the central charges
transform as a doublet under this group. There is also a pair of
grading operators $(d_1, d_2)$ which assign the weight $(n,m)$ to the
generator $W_{n,m}$.

We are not going to write down the relations for the DIM algebra for
$t \neq q$, since they can be easily found in the
literature~\cite{DIM}. There are, however, two general remarks. First
of all, the $t$-deformed algebra is still $SL(2,
\mathbb{Z})$-invariant. Secondly, the algebra is symmetric under the
permutation of the parameters $q$, $t^{-1}$ and $\frac{t}{q}$. The
remnant of this large symmetry is the symmetry of Eq.~\eqref{eq:31}
under $q \leftrightarrow q^{-1}$ (one needs to simultaneously invert
the sign of $W_{n,m}$).

We will describe the MacMahon representation $M_c(u)$ in the basis
corresponding to the differential operators $w_{n,m} \simeq z^n
(z \partial_z)^m$, instead of the difference operators $W_{n,m} \simeq
z^n q^{m z\partial_z}$ used in Eq.~\eqref{eq:31}, but the argument is
similar in both cases. Notice the difference between the indices in
$w_{n,m}$ and $W_{n,m}$: in the first case $m \in \mathbb{Z}_{\geq
  0}$, while in the second $n$ and $m$ are general integers.

Consider the highest weight representation with central charge $(c_1,
c_2) = (c,0) $. The highest weight state $|v\rangle$, is the
eigenstate of $w_{0,m}$ annihilated by $w_{n,m}$ for $n \geq 1$. The
states of the representation are obtained by acting on $|v\rangle$
with the operators $w_{n,m}$ with $n <0$ and $m\geq 0$.

We further require there to be only a finite number of states on a
given level. This is an extremely stringent requirement which fixes
the allowed eigenvalues of $w_{0,m}$ almost completely. To get a
finite number of states on each level we need to have a lot of
null-states. It turns out that the simplest nontrivial representation
of this form has the null-subspaces generated by the following states:
\begin{equation}
  \label{eq:6}
  \text{Null states:}\qquad |\chi_n\rangle \simeq z^{-n} \prod_{j=0}^{n-1}(z \partial_z - j)
  |v\rangle \simeq \partial_z^n |v \rangle, \qquad n > 0
\end{equation}
where we have used the identification between the generators and the
differential operators to write the null-states explicitly. Following
the pattern of null states, we see that there are exactly $n$
independent generators $w_{-n,m}$ for a given degree $n > 0$: those
for which $0\leq m \leq n-1$. All the other generators produce
null-states. This situation is exactly parallel to the chiral free
boson Hilbert space $\mathcal{H}^{3d}$, which is generated by
$a_{-(j,m)}$ with $1 \leq m \leq j$. Thus, there is a direct
equivalence between the holomorphic states of the $3d$ free
boson~\eqref{eq:30} and the states of the MacMahon representation
$M_c(u)$ of $W_{1+\infty}$:
\begin{equation}
  \label{eq:7}
  \begin{array}{c|c}
    W_{1+\infty} & 3d \text{ boson} \\
    \hline
    w_{-n_1, m_1} w_{-n_2, m_2}\cdots w_{-n_k, m_k} |v\rangle &
    a_{-(n_1, m_1+1)} a_{-(n_2, m_2+1)}\cdots a_{-(n_k, m_k+1)}
    |\varnothing \rangle\\
    n_i > 0,\quad 0 \leq m_i \leq n_i-1 &     n_i > 0,\quad 1 \leq m_i+1 \leq n_i
  \end{array}
\end{equation}

The only essential difference between the two pictures is that the
creation and annihilation operators $a_{-(j,m)}$ with non-opposite
vectors $(j,m)$ \emph{commute} while the commutator of two generators
$W_{n,m}$ is nonzero in general.

Let us look for the analogy to this situation in the $2d$ free
boson. Indeed, there are creation operators $a_{-m}$, which commute
for non-opposite $m$ and there are also \emph{Virasoro} generators
$L_{-n} \sim \sum_{k\in \mathbb{Z}} \normord{a_{k-n} a_{-k}}$, which
form a nontrivial Lie algebra.  We can hypothesize that the same
phenomenon happens in the $3d$ case and the $W_{1+\infty}$ generators
are expressed as \emph{bilinear combinations} of bosonic operators
$a_{(j,m)}$:
\begin{equation}
  \label{eq:8}
  w_{n,m} \stackrel{?}{\sim} \sum_{k,l} \normord{a_{n+k,m+l} a_{-k,-l}}.
\end{equation}
However, before applying Eq.~\eqref{eq:8} one needs to understand the
limits of summation, the precise nature of the normal ordering and
possible convergence issues. We will not attempt this task here.

\section{Affine matrix models and DIM Bethe equations}
\label{sec:affine-matrix-models}
In this section we derive the analogue of the quantum spectral curve
for affine matrix models. Taking its Nekrasov-Shatashvili limit we
obtain Baxter and Bethe equations, which are slight generalizations of
those given in~\cite{ILW, FJMM}. Their solutions are labelled by plane
partitions. Following the results of the previous section this hints
at a possible connection with an \emph{integrable} $3d$ field theory.

Affine matrix model is one of the species from the zoo of ``network
matrix models''~\cite{MMZ} obtained from refined topological
strings~\cite{ref}. Concretely, this model corresponds to the
compactified strip geometry. In the geometric engineering terms it
gives the $5d$ $\mathcal{N}=1$ $U(K)$ gauge theory with an extra
adjoint multiplet, or equivalently $6d$ abelian linear quiver. The
matrix model average of a function $f(\vec{x})$ is given by:
\begin{equation}
  \label{eq:32}
  \langle f(\vec{x}) \rangle_{N,u,\vec{z},\vec{w}} = \frac{1}{Z} \oint_{\mathcal{C}} \prod_{i=1}^N \left\{ \frac{dx_i}{x_i} x_i^u \prod_{a=1}^K \frac{\left(
        \frac{w_a}{x_i};q \right)_{\infty}}{\left( \frac{z_a}{x_i};q
      \right)_{\infty}} \right\} \prod_{i \neq j}^N \frac{\left( \frac{x_i}{x_j};
      q\right)_{\infty} \left(t_1 t_2 \frac{x_i}{x_j};
      q\right)_{\infty}}{\left( t_1 \frac{x_i}{x_j};
      q\right)_{\infty} \left( t_2 \frac{x_i}{x_j};
      q\right)_{\infty}} f(\vec{x}),
\end{equation}
where $Z$ is the integral without the insertion and the contour
$\mathcal{C}$ can be chosen to encircle the poles of the
integrand. These poles correspond to $K$-tuples of plane partitions
$\pi^{(a)}$, $a=1,\ldots,K$:
\begin{equation}
  \label{eq:33}
  x_I = z_a q^{\pi^{(a)}_{i,j}-1} t_1^{1-i} t_2^{1-j}. 
\end{equation}
The partitions $\pi^{(a)}$ have the total \emph{floor area} $N$, while
the total number of boxes can be arbitrary. The residues at the poles
transform naturally under arbitrary permutations of $t_1$, $t_2$ and
$q^{-1}$ and the simultaneous transpositions of the plane partitions
$\pi^{(a)}$ (assuming $N$ to be sufficiently large).

\subsection{Quantum spectral curve}
\label{sec:quant-spectr-curve}
We follow the standard technique to obtain the quantum spectral curve
(or loop equations, or Ward identities, or $qq$-characters) for the
model and consider an integral of a total difference\footnote{The
  contour integral in the definition of the model is equivalent to a
  Jackson integral, for which total differences play the role of total
  derivatives.}:
\begin{equation}
  \label{eq:9}
  0 = \oint_{\mathcal{C}} \prod_{k=1}^N \frac{dx_k}{x_k} \sum_{i=1}^N
  (1 - q^{x_i \partial_{x_i}}) \left\{
  \frac{x_i^3  \prod_{a=1}^K(x_i-z_a)}{x_i - \xi} \prod_{j \neq i}
  \frac{\left( \frac{t_1 t_2}{q} x_i - x_j \right) \left( t_1^{-1} x_i
      - x_j \right) \left( t_2^{-1} x_i - x_j \right)}{(x_i - x_j)}
  \mu(\vec{x}) \right\}, 
\end{equation}
where
\begin{equation}
  \label{eq:10}
  \mu(\vec{x}) = \prod_{i=1}^N \left\{  x_i^u \prod_{a=1}^K \frac{\left(
        \frac{w_a}{x_i};q \right)_{\infty}}{\left( \frac{z_a}{x_i};q
      \right)_{\infty}} \right\} \prod_{i \neq j}^N \frac{\left( \frac{x_i}{x_j};
      q\right)_{\infty} \left(t_1 t_2 \frac{x_i}{x_j};
      q\right)_{\infty}}{\left( t_1 \frac{x_i}{x_j};
      q\right)_{\infty} \left( t_2 \frac{x_i}{x_j};
      q\right)_{\infty}}
\end{equation}
is the matrix model measure. The difference operator acts on the
measure as follows:
\begin{equation}
  \label{eq:11}
  q^{x_i \partial_{x_i}} \mu(\vec{x}) = q^u \prod_{a=1}^K \frac{q x_i -
    w_a}{q x_i - z_a} \prod_{i\neq j}
  \frac{(q x_i - x_j) ( t_1 x_i - x_j ) ( t_2 x_i
      - x_j ) \left( \frac{q}{t_1 t_2} x_i - x_j \right)}{(x_i - x_j) ( t_1 t_2 x_i - x_j ) \left( \frac{q}{t_1} x_i
      - x_j \right) \left( \frac{q}{t_2} x_i - x_j \right)} \mu(\vec{x})
\end{equation}
so that half of the factors cancel with the extra factors in
Eq.~\eqref{eq:9}. The identity for the averages following from
Eq.~\eqref{eq:9} is
\begin{multline}
  \label{eq:34}
  \Biggl\langle \sum_{i=1}^N \frac{x_i^3 \prod_{a=1}^K(x_i-z_a)}{x_i -
    \xi} \prod_{j \neq i} \frac{\left( \frac{t_1 t_2}{q} x_i - x_j
    \right) \left( t_1^{-1} x_i - x_j \right) \left( t_2^{-1} x_i -
      x_j \right)}{(x_i - x_j)}
  -\\
  - \sum_{i=1}^N q^{u+3} \frac{x_i^3 \prod_{a=1}^K (q x_i - w_a)}{q x_i - \xi}
  \prod_{j\neq i} \frac{( t_1 x_i - x_j ) ( t_2 x_i - x_j ) \left(
      \frac{q}{t_1 t_2} x_i - x_j \right)}{(x_i - x_j)} \Biggr\rangle =
  0.
\end{multline}
We can use one more standard trick~\cite{Zenkevich:2014lca, Z2} and
rewrite the sums under the average as contour integrals over an
auxiliary parameter $y$:
\begin{multline}
  \label{eq:35}
  \sum_{i=1}^N \frac{x_i^3 \prod_{a=1}^K(x_i-z_a)}{x_i - \xi}
  \prod_{j \neq i} \frac{\left( \frac{t_1 t_2}{q} x_i - x_j \right)
    \left( t_1^{-1} x_i - x_j \right) \left( t_2^{-1} x_i - x_j
    \right)}{(x_i - x_j)}
  =\\
  = \frac{1}{\left( \frac{t_1 t_2}{q} - 1 \right) (t_1^{-1} -
    1)(t_2^{-1} - 1)} \oint_{\mathcal{C}_x} dy
  \frac{\prod_{a=1}^K(y-z_a)}{(y - \xi)}
  \prod_{j = 1}^N \frac{\left( \frac{t_1 t_2}{q} y - x_j \right)
    \left( t_1^{-1} y - x_j \right) \left( t_2^{-1} y - x_j
    \right)}{(y - x_j)}
\end{multline}
\begin{multline}
  \label{eq:36}
    \sum_{i=1}^N \frac{x_i^3 \prod_{a=1}^K(qx_i-w_a)}{q x_i - \xi}
  \prod_{j \neq i} \frac{\left( \frac{q}{t_1 t_2} x_i - x_j \right) \left( t_1 x_i - x_j \right) \left( t_2 x_i - x_j
    \right) }{(x_i - x_j)}
  =\\
  = \frac{1}{\left( \frac{q}{t_1 t_2} - 1 \right) (t_1 -
    1)(t_2 - 1)} \oint_{\mathcal{C}_x} dy
  \frac{\prod_{a=1}^K(q y - w_a)}{(q y - \xi)}
  \prod_{j = 1}^N \frac{\left( \frac{q}{t_1 t_2} y - x_j \right)
    \left( t_1 y - x_j \right) \left( t_2 y - x_j
    \right) }{(y - x_j)}
\end{multline}
where the contour $\mathcal{C}_x$ encircles all the points
$x_j$. Deforming the contour we pick up the residues at $y=\xi$ and at
$y=0, \infty$. The former residue gives the expression for the quantum
spectral curve while the latter two produce polynomials in $\xi$ with
coefficients polynomially depending on $x_i$.
\begin{equation}
  \label{eq:37}
\boxed{    K_{+}(\xi) \left\langle \frac{Q \left( \frac{t_1
          t_2}{q} \xi \right) Q \left( \frac{\xi}{t_1} \right) Q
      \left( \frac{\xi}{t_2} \right)}{Q \left(\xi \right)} \right\rangle +q^{u+1}
    K_{-}(\xi) \left\langle \frac{Q \left( \frac{\xi}{t_1
          t_2} \right) Q \left( \frac{t_1}{q} \xi \right) Q \left(
        \frac{t_2}{q} \xi \right)}{Q \left(\frac{\xi}{q} \right)}
  \right\rangle = \mathrm{Pol}_{2N+K}(\xi)}
\end{equation}
where
\begin{equation}
  \label{eq:39}
  Q(\xi) = \prod_{j=1}^N (\xi - x_j),\qquad K_{+}(\xi) =
  \prod_{a=1}^K(\xi-z_a),\qquad K_{-}(\xi) =
  \prod_{a=1}^K(\xi-w_a).
\end{equation}
This equation can be viewed as a three-parametric ($t_1$, $t_2$ and
$q$) \emph{deformation} of the Baxter equation associated to DIM
algebra~\cite{FJMM}. Notice also that the equation is symmetric under
permutation of $t_1$, $t_2$ and $\frac{q}{t_1 t_2}$. This symmetry
combines with the symmetry of the integrand under the permutation of
$t_1$, $t_2$ and $q^{-1}$, which we mentioned in the beginning of this
section. Overall, the matrix model is symmetric under permutation of
\emph{four parameters,} $t_1$, $t_2$, $q^{-1}$ and $\frac{q}{t_1 t_2}$
with their product being equal to one.

\subsection{NS limit and Bethe equations}
\label{sec:ns-limit-bethe}
In the NS limit $q \to 1$ the matrix model integral can be evaluated
by saddle point method. The quantum spectral curve
equation~\eqref{eq:37} is then understood as an equation determining
the saddle points $\vec{x}$. The average signs in Eq.~\eqref{eq:37},
therefore, disappear and it reduces to the DIM Baxter TQ equation:
\begin{equation}
  \label{eq:44}
  K_{+}(\xi) \frac{Q \left( \frac{t_1
          t_2}{q} \xi \right) Q \left( \frac{\xi}{t_1} \right) Q
      \left( \frac{\xi}{t_2} \right)}{Q \left(\xi \right)}  +q^{u+1}
  K_{-}(\xi) \frac{Q \left( \frac{\xi}{t_1
          t_2} \right) Q \left( \frac{t_1}{q} \xi \right) Q \left(
        \frac{t_2}{q} \xi \right)}{Q \left(\frac{\xi}{q} \right)}
   = \mathrm{Pol}_{2N+K}(\xi).
 \end{equation}
 where $Q(\xi)$ is a polynomial with roots $x_i$. The only difference
 between our equation and the standard one considered in the
 literature~\cite{FJMM} is that the roots of the polynomials
 $K_{+}(\xi)$ and $K_{-}(\xi)$ are \emph{a priori} not related to each
 other. In~\cite{FJMM}, however, one had the condition $K_{-}(\xi) =
 K_{+}(t_1 t_2 \xi)$. In the matrix model language this constraint
 relating $w_a$ and $z_a$ is interpreted as the \emph{degeneracy
   condition} for the vertex operators sitting at points $z_a$ ---
 their weights (dimensions) take special, quantized values. As we will
 see momentarily, lifting of the constraint has dramatic consequences
 for the structure of the solutions: \emph{with} the constraint they
 are enumerated (up to permutations) by $K$-tuples of Young diagrams,
 while \emph{without} it the solutions are much more numerous and
 correspond to $K$-tuples of plane partitions.

Bethe equations following from the Baxter equation~\eqref{eq:44} are:
\begin{equation}
  \label{eq:38}
  K_{+}(x_i) \prod_{j\neq i} \frac{(t_1 t_2 x_i - x_j) \left(
      \frac{x_i}{t_1} - x_j \right) \left( \frac{x_i}{t_2} - x_j
    \right)}{x_i - x_j} =   e^{\tau} K_{-}(x_i) \prod_{j\neq i} \frac{\left(\frac{x_i}{t_1 t_2} - x_j\right) \left(
      t_1 x_i - x_j \right) \left( t_2 x_i - x_j
    \right)}{x_i - x_j}
\end{equation}
where $e^{\tau} = \lim_{q \to 1} q^{u+1}$ (we assume that $u$ scales
as $\frac{1}{\ln q}$ in the NS limit). Let us consider two limits
$\tau \to \pm \infty$ of these equations and see how plane partitions
arise as solutions. Our analysis is very similar to that
of~\cite{ILW}, though the end results are different for the reason
discussed above.

\begin{enumerate}
\item $\tau \to - \infty$. Equations~\eqref{eq:38} simplify into:
\begin{equation}
  \label{eq:40}
  K_{+}(x_i) \prod_{j\neq i} \frac{(t_1 t_2 x_i - x_j) \left(
      \frac{x_i}{t_1} - x_j \right) \left( \frac{x_i}{t_2} - x_j
    \right)}{x_i - x_j} = 0.
\end{equation}
Suppose for a moment that $K=1$, $N=2$ and one of the variables, say
$x_1$ sits at point $z_1$. Then, equation~\eqref{eq:40} for $i=1$ is
already satisfied and we need not consider it anymore. The next
variable, $x_2$ should solve one of the remaining equations. It cannot
sit at $z_1$ because of the denominator $(x_1 - x_2)$, and therefore
should be either $t_1z_1$, $t_2z_1$ or $(t_1 t_2)^{-1}z_1$. These
three solutions correspond to three plane partitions with $N=2$ boxes.

In general one can see that the solutions are labelled by $K$-tuples
of plane partitions $\pi'^{(a)}$ with total of $N$ boxes. They are
given by
\begin{equation}
  \label{eq:41}
  x_i = z_a  t_1^{1-i} t_2^{1-j} (t_1 t_2)^{k-1}, \qquad (i,j,k)\in \pi'^{(a)}.
\end{equation}
We use a prime to distinguish partitions $\pi'^{(a)}$ from the
partitions $\pi^{(a)}$ from Eq.~\eqref{eq:33} labelling the poles of
the affine matrix model integrand. In the NS limit the poles condense,
so that the size of the saddle point partitions $\pi^{(a)}$ is
actually infinite. Notice that $N$ is the \emph{total number of boxes}
in $\pi'^{(a)}$, while the partitions $\pi^{(a)}$ have total
\emph{floor area} $N$ and arbitrary number of boxes.

\item $\tau \to +\infty$. This limit is analogous to the previous one:
  \begin{equation}
    \label{eq:42}
    K_{-}(x_i) \prod_{j\neq i} \frac{\left(\frac{x_i}{t_1 t_2} - x_j\right) \left(
        t_1 x_i - x_j \right) \left( t_2 x_i - x_j
      \right)}{x_i - x_j} = 0.
  \end{equation}
  The solutions are again labelled by plane partitions, but the roots
  cluster around the points $w_a$, not~$z_a$:
  \begin{equation}
    \label{eq:43}
    x_i = w_a  t_1^{i-1} t_2^{j-1} (t_1 t_2)^{1-k}, \qquad (i,j,k)\in \pi'^{(a)}.
  \end{equation}
\end{enumerate}

The solutions for general $\tau$ extrapolate between two limits $\tau
\to \pm \infty$, but are still enumerated by plane partitions. We have
verified this claim by numerically solving the equations for small $N$
and $K$. We can then ask what happens in the limit when $w_a \to t_1
t_2 z_a$? One can notice that in this case plane partitions with
height more than one cease to be solutions, because of extra
cancellations between the factors in $K_{+}$ and $K_{-}$. We have also
verified this effect numerically.

The solutions to Bethe equations correspond to the basis of
eigenstates of DIM integrals of motion in the corresponding DIM
representation. For general $K_{+}$ and $K_{-}$ the representation is
a tensor product of MacMahon modules $M_{\frac{w_1}{z_1}}(z_1) \otimes
\cdots \otimes M_{\frac{w_K}{z_K}}(z_K)$ with spectral parameters
$z_a$ and central charges $(\ln \frac{w_a}{z_a}, 0)$. The basis is
given by the $K$-tuples of plane partitions. For special degenerate
value of the central charge equal to $(\ln (t_1 t_2), 0)$ the MacMahon
module becomes reducible. After factoring out the invariant subspace
one obtains the \emph{Fock} representation with standard basis
labelled by Young diagrams. The disappearing solutions of the Bethe
equations correspond to the invariant subspaces factored out from the
MacMahon modules.

Let us summarize the results of this section. We have derived the
quantum spectral curve equation for the affine matrix model and
studied the Bethe equations obtained in the NS limit. The solutions of
these equations are in general labelled by $K$-tuples of plane
partitions and span the tensor product of MacMahon representations of
DIM algebra $U_{t_1, t_2} (\widehat{\widehat{\mathfrak{gl}}}_1)$. The
Bethe equations, therefore describe certain (eigen)states of a $3d$
field field theory. In the next section we give an explicit
construction of the basis of eigenstates in the single MacMahon
module.

\section{Triple Macdonald polynomials}
\label{sec:triple-macd-polyn}
In this section we build the states corresponding to the solutions of
Bethe equations~\eqref{eq:40} for the simplest case of $K=1$. We model
the MacMahon module as a subspace inside the infinite tensor product
of Fock modules $\mathcal{F}(u)$ of DIM algebra $U_{t_1,t_2}
(\widehat{\widehat{\mathfrak{gl}}}_1)$ with specially adjusted
spectral parameters. The infinite tensor product of Fock spaces, each
of which is effectively a $2d$ free boson, can be seen as a
discretized construction of a $3d$ field.

Concretely, we consider the eigenfunctions of the Cartan subalgebra of
DIM algebra. The first element of this subalgebra is the zero mode of
the current $x_0^{+}$, which corresponds to generator $W_{0,1}$ in the
$W_{1+\infty}$ notation. We will diagonalize this element in the
tensor product of Fock representations\footnote{The definition of the
  tensor product of representations requires the choice of the
  coproduct $\Delta$ in the DIM algebra. This amounts to choosing a
  preferred direction in the space of central charge vectors. Here we
  choose the horizontal direction.} $\mathcal{F}(u_1) \otimes \cdots
\otimes \mathcal{F}(u_L)$, which we identify with the space of
polynomials in $L$ families of time variables $p_{a,n}$, $a=1,\ldots,
L$, $n \geq 1$. The action of $x_0^{+}$ on this space is written as
follows:
\begin{equation}
  \label{eq:46}
  x_0^{+}|_{\mathcal{F}(u_1) \otimes \cdots
\otimes \mathcal{F}(u_L)} = \oint_{\mathcal{C}_0} \frac{dz}{z}
  \sum_{a=1}^L u_a \Lambda_a(z),
\end{equation}
where $\mathcal{C}_0$ is a small contour around zero and
\begin{equation}
  \label{eq:47}
  \Lambda_a(z) = \exp \left\{ \sum_{n \geq 1} \frac{1-t_2^n}{n} z^n
    \left[p_{a,n}  + \left( 1 - (t_1 t_2)^{-n} \right) \sum_{b=1}^{a-1} p_{b,n}
    \right] \right\} \exp \left\{ \sum_{n \geq 1} (1-t_1^n) z^{-n}
    \frac{\partial}{\partial p_{a,n}}
  \right\}.
\end{equation}
The equation for the eigenfunctions reads
\begin{equation}
  \label{eq:48}
  x_0^{+}|_{\mathcal{F}(u_1) \otimes \cdots
\otimes \mathcal{F}(u_L)} M_{\vec{Y}}^{(t_1,t_2)}(\vec{u}|p_{a,n}) =
\sum_{a=1}^L u_a \left[1-  (1-t_1)(1-t_2) \sum_{(i,j)\in Y_a} t_1^{j-1}
  t_2^{i-1}  \right] M_{\vec{Y}}^{(t_1,t_2)}(\vec{u}|p_{a,n})
\end{equation}
The eigenfunctions $M_{\vec{Y}}^{(t_1, t_2)}(\vec{u}|p_{a,n})$ are
called generalized Macdonald
polynomials~\cite{Zenkevich:2014lca},~\cite{GenMD} and depend on
$L$-tuple of Young diagrams $\vec{Y}$ and $L$-tuple\footnote{ More
  precisely $M_{\vec{Y}}^{(t_1,t_2)}(\vec{u}|p_{a,n})$ depend only on
  $(L-1)$ \emph{ratios} of $u_a$.} of spectral parameters $\vec{u}$.

For certain special values of spectral parameters called the
resonances, i.e.\ when $u_a = u_{a+1} t_1^m t_2^l$, part of the tensor
product of Fock modules completely decouples (i.e.\ the corresponding
states become null-vectors inside the DIM representation) leaving only
polynomials corresponding to the \emph{special} $L$-tuples of Young
diagrams.  The same phenomenon can be observed when considering the
Nekrasov expansion of the conformal block. To get the right answer for
\emph{degenerate} values of the intermediate dimensions one should
constraint the Young diagrams appearing in the expansion to satisfy
the \emph{Burge conditions}~\cite{Bershtein:2014qma}. We would like to
consider here the simplest case of the resonance when $u_a = u (t_1
t_2)^{1-a}$. In this case the surviving $L$-tuples of Young diagrams
form a plane partition $\pi = \{ Y_1, \ldots , Y_L\}$ of height~$L$.

Another way to see the decoupling of states for spectral parameters in
resonance is to consider the $R$-matrix for the DIM
algebra~\cite{DIMR} acting in the tensor product of Fock spaces
$\mathcal{F}(u) \otimes \mathcal{F}(\frac{u}{x})$. The $R$-matrix in
the resonant case becomes singular. Indeed, the eigenvalues of the
$R$-matrix can be read off from its expression in the vertical
representation (see~\cite{DIMR} for details):
\begin{equation}
  \label{eq:5}
  \mathcal{R}_{\lambda \mu}(x) = \left( t_1 t_2
  \right)^{\frac{|\lambda| + |\mu|}{2}} \frac{G_{\lambda
      \mu}^{(t_1,t_2^{-2})}(x)}{G_{\lambda \mu}^{(t_1,t_2^{-1})}\left(
      t_1 t_2 x \right)}.
\end{equation}
For $x= t_1 t_2$ the denominator in Eq.~\eqref{eq:5} is finite for any
Young diagrams $\lambda$ and $\mu$ while the numerator vanishes
whenever $\mu$ doesn't fit inside $\lambda$. Since the operator
$x_0^{+}$ is an element of DIM algebra, its action on the tensor
product commutes with the $R$-matrix. Thus, using the $R$-matrix one
can project out all the states $|\lambda\rangle \otimes |\mu\rangle$
in the tensor product $\mathcal{F}(u) \otimes \mathcal{F}(\frac{t_1
  t_2}{x})$, except those satisfying $\lambda \supset \mu$.

The effect here is similar to the fusion of spins in a spin
chain. Take as an example the rational $R$-matrix $R(x-y) = 1 -
\frac{P}{x-y}$ acting in the tensor product of two fundamental
evaluation representations $\mathbb{C}^n$ at points $x$ and $y$ ($P$
is the permutation operator). This $R$-matrix becomes a projector for
the resonant values of the spectral parameters $x = y \pm 1$. For
these values one can take a projection of the tensor product using
$R(\pm 1)$ and, iterating this procedure, build evaluation
representations with general spin.

Back to the DIM case: we notice that generalized Macdonald polynomials
are \emph{stable} in the following sense. Consider an $L$-tuple of
Young diagrams $\vec{Y}$, in which only first $m$ diagrams are
non-empty. Then $M_{\vec{Y}}^{(t_1,t_2)}(\vec{u}|p_{a,n})$ does not
depend on $L$ as long as $L \geq m$. In particular,
$M_{\vec{Y}}^{(t_1,t_2)}(\vec{u}|p_{a,n})$ in this case depends only
on the first $m$ times $p_{a,n}$, $a=1,\ldots ,m$ but not on those
with $m < a \leq L$. Using this stability property we can take the
limit $L\to \infty$ of the polynomials $M_{\vec{Y}}^{(t_1,t_2)}(u (t_1
t_2)^{1-a}|p_{a,n})$. We denote the resulting polynomials by $M^{(t_1,
  t_2, (t_1 t_2)^{-1})}_{\pi} (p)$ and call them \emph{triple}
Macdonald polynomials. They depend on the plane partition $\pi$ and an
infinite number of families of time variables $p_{a,n}$ $a \geq 1$,
though for any concrete $\pi$ only a finite number of $p_{a,n}$ enters
$M^{(t_1, t_2, (t_1 t_2)^{-1})}_{\pi} (p)$.

The eigen-equation for triple Macdonald polynomials reads:
\begin{multline}
  \label{eq:52}
 \left(1 - (t_1 t_2)^{-1}\right) x_0^{+}|_{\bigotimes_{a \geq 1} \mathcal{F}(u (t_1 t_2)^{1-a})}
  M^{(t_1, t_2, (t_1 t_2)^{-1})}_{\pi} (p) =\\
  = u \left[ 1 - (1-t_1)(1-t_2) \left(1 - (t_1 t_2)^{-1}\right)
    \sum_{(i,j,a)\in \pi} t_1^{j-1} t_2^{i-1} (t_1 t_2)^{1-a}
  \right] M^{(t_1, t_2, (t_1 t_2)^{-1})}_{\pi} (p)
\end{multline}
Notice how the eigenvalues in the r.h.s.\ of Eq.~\eqref{eq:52} can be
expressed as the sum of the Bethe roots $x_i$ from
Eq.~\eqref{eq:43}. The eigenvalues are invariant with respect to the
permutation of $t_1$, $t_2$ and $(t_1 t_2)^{-1}$ and the corresponding
simultaneous transposition of $\pi$. Only part of this symmetry
exchanging $t_1$ and $t_2$ is manifest in our description, and it
would be very interesting to understand the other hidden part.

The most elementary property of Macdonald polynomials is that for $t_1
= t_2^{-1}$ they reduce to the product of Schur polynomials:
\begin{equation}
  \label{eq:49}
   M^{(t_1, t_1^{-1}, 1)}_{\pi} (p) = \prod_{i=1}^{h(\pi)} s_{\pi^{(i)}}(p_{i,n})
\end{equation}
They also satisfy the inversion relation analogous to that of
ordinary Macdonald polynomials:
\begin{equation}
  \label{eq:50}
  M^{(t_1, t_2, (t_1 t_2)^{-1})}_{\pi} \left( - \frac{1 - t_1^n}{1 -
      t_2^{-n}} p_{a,n}\right) = (-1)^{|\pi|} \prod_{i=1}^{h(\pi)}
  \frac{C_{\pi^{(i)}}}{C_{\pi^{(i)}}'}   M^{(t_2, t_1, (t_1
    t_2)^{-1})}_{\pi^{\mathrm{T}}} (p)
\end{equation}
where $\pi^{\mathrm{T}}$ denotes one of the transpositions of $\pi$,
which acts on the slices of fixed height, and
\begin{equation}
  \label{eq:51}
  C_Y = \prod_{(i,j)\in Y} (1 - t_1^{Y_i - j +1} t_2^{i -
    Y_j^{\mathrm{T}}}), \qquad   C_Y' = \prod_{(i,j)\in Y} (1 - t_1^{Y_i - j} t_2^{i -
    Y_j^{\mathrm{T}}-1}).
\end{equation}

\subsection{Examples}
\label{sec:examples}
Let us give explicit expressions for triple Macdonald polynomials on
the first three levels:
\begin{align}
  \label{eq:45}
  M_{[[1]]}^{(t_1,t_2, (t_1 t_2)^{-1})}(p)&= p_{1,1}\notag\\
  M_{[[2]]}^{(t_1,t_2, (t_1 t_2)^{-1})}(p)&=\frac{\left(t_1-1\right) \left(t_2+1\right) p_{1,2}}{2
   \left(t_1-t_2\right)}-\frac{\left(t_1+1\right) \left(t_2-1\right)
   p_{1,1}^2}{2 \left(t_1-t_2\right)}\notag\\
  M_{[[1,1]]}^{(t_1,t_2, (t_1 t_2)^{-1})}(p)&=\frac{p_{1,1}^2}{2}-\frac{p_{1,2}}{2}\notag\\
  M_{[[1],[1]]}^{(t_1,t_2, (t_1 t_2)^{-1})}(p)&=-\frac{\left(t_1 t_2-1\right) \left(t_2^2 t_1^2+t_2 t_1^2+t_2^2 t_1-t_2
   t_1-2\right) p_{1,1}^2}{2 t_1 t_2 \left(t_1^2 t_2-1\right) \left(t_1
   t_2^2-1\right)}+\frac{\left(t_1-1\right) \left(t_2+1\right) \left(t_1
   t_2-1\right) p_{1,2}}{2 \left(t_1^2 t_2-1\right) \left(t_1
   t_2^2-1\right)}+p_{2,1} p_{1,1}\notag
\end{align}
\begin{align}
M_{[[3]]}^{(t_1,t_2, (t_1 t_2)^{-1})}(p)&= \frac{\left(t_1+1\right) \left(t_1^2+t_1+1\right) \left(t_2-1\right){}^2
   p_{1,1}^3}{6 \left(t_1-t_2\right)
   \left(t_1^2-t_2\right)}-\frac{\left(t_1-1\right) \left(t_1^2+t_1+1\right)
   \left(t_2-1\right) \left(t_2+1\right) p_{1,2} p_{1,1}}{2 \left(t_1-t_2\right)
   \left(t_1^2-t_2\right)}+\notag\\
 &\phantom{=}+\frac{\left(t_1-1\right){}^2 \left(t_1+1\right)
   \left(t_2^2+t_2+1\right) p_{1,3}}{3 \left(t_1-t_2\right)
   \left(t_1^2-t_2\right)}\notag\\
M_{[[2,1]]}^{(t_1,t_2, (t_1 t_2)^{-1})}(p)&= -\frac{\left(t_2-1\right) \left(t_2 t_1+2 t_1+2 t_2+1\right) p_{1,1}^3}{6
   \left(t_1-t_2^2\right)}+\frac{\left(t_2+1\right) \left(t_1 t_2-1\right)
   p_{1,2} p_{1,1}}{2 \left(t_1-t_2^2\right)}-\frac{\left(t_1-1\right)
   \left(t_2^2+t_2+1\right) p_{1,3}}{3 \left(t_1-t_2^2\right)}\notag\\
M_{[[1,1,1]]}^{(t_1,t_2, (t_1 t_2)^{-1})}(p)&= \frac{p_{1,1}^3}{6}-\frac{1}{2} p_{1,2} p_{1,1}+\frac{p_{1,3}}{3}\notag\\
  M_{[[2],[1]]}^{(t_1,t_2, (t_1 t_2)^{-1})}(p)&= \frac{\left(t_1+1\right) \left(t_2-1\right) \left(t_1 t_2-1\right) \left(t_2^2
   t_1^3+2 t_2 t_1^3+t_2^2 t_1^2-t_2 t_1^2+t_2^2 t_1-t_2 t_1-3\right)
   p_{1,1}^3}{6 t_1 \left(t_1-t_2\right) t_2 \left(t_1^3 t_2-1\right) \left(t_1
   t_2^2-1\right)}-\notag\\
&\phantom{=} \frac{\left(t_1+1\right) \left(t_2-1\right) p_{2,1}
 p_{1,1}^2}{2 \left(t_1-t_2\right)}- \frac{\left(t_1-1\right)
   \left(t_2+1\right) \left(t_1 t_2-1\right) \left(t_2^2 t_1^3+t_2^2 t_1^2-t_2
   t_1^2+t_2^2 t_1-t_2 t_1-1\right) p_{1,2} p_{1,1}}{2 t_1 \left(t_1-t_2\right)
   t_2 \left(t_1^3 t_2-1\right) \left(t_1
   t_2^2-1\right)}+\notag\\
&\phantom{=}+\frac{\left(t_1-1\right){}^2 \left(t_1+1\right) \left(t_1
   t_2-1\right) \left(t_2^2+t_2+1\right) p_{1,3}}{3 \left(t_1-t_2\right)
 \left(t_1^3 t_2-1\right) \left(t_1 t_2^2-1\right)}+\frac{\left(t_1-1\right)
   \left(t_2+1\right) p_{1,2} p_{2,1}}{2 \left(t_1-t_2\right)} \notag\\
M_{[[1,1],[1]]}^{(t_1,t_2, (t_1 t_2)^{-1})}(p)&= -\frac{\left(t_1 t_2-1\right) \left(t_1^2 t_2^3+2 t_1 t_2^3+t_1^2 t_2^2-t_1
   t_2^2+t_1^2 t_2-t_1 t_2-3\right) p_{1,1}^3}{6 t_1 t_2 \left(t_1^2
   t_2-1\right) \left(t_1 t_2^3-1\right)}+\notag\\
&\phantom{=}+\frac{\left(t_1 t_2-1\right)
   \left(t_1^2 t_2^3+t_1^2 t_2^2-t_1 t_2^2+t_1^2 t_2-t_1 t_2-1\right) p_{1,2}
   p_{1,1}}{2 t_1 t_2 \left(t_1^2 t_2-1\right) \left(t_1
   t_2^3-1\right)}-\notag\\
&\phantom{=}-\frac{\left(t_1-1\right) \left(t_1 t_2-1\right)
   \left(t_2^2+t_2+1\right) p_{1,3}}{3 \left(t_1^2 t_2-1\right) \left(t_1
   t_2^3-1\right)}+\frac{1}{2} p_{2,1} p_{1,1}^2-\frac{1}{2} p_{1,2} p_{2,1} \notag
\end{align}

\section{Conclusions and outlook}
\label{sec:conclusions}

We have investigated several situations in which plane partitions
naturally arise. Technically, they label the eigenvectors inside the
MacMahon representation of DIM algebra and the corresponding solutions
to Bethe equations. We give arguments that the natural framework to
work with such representations is not the $2d$ fields, but $3d$ ones,
where the (chiral) Hilbert space is spanned by plane partitions.

Let us give directions in which we would like to extend our results in
the future.

Matrix models of the type we considered can be understood as
Dotsenko-Fateev representations~\cite{DF} of certain $W$-algebra
conformal blocks. The quantum spectral curve~\eqref{eq:37} corresponds
to a particular generator of the $W$-algebra commuting with a set of
screening charges, whose correlator gives the affine matrix
model. This $W$-algebra is associated to a circular quiver with one
node~\cite{Kimura:2015rgi}. The advantage of our quantum spectral
curve is that it is a finite expression, whereas the generators given
in~\cite{Kimura:2015rgi} were infinite series. One can notice that the
origin of this difference lies in the difference between classes of
DIM representations, since the $W$-algebra provides the quantization
of DIM representation ring. The representation corresponding
to~\eqref{eq:37} is what is called in~\cite{FJMM} a module of ``finite
type'', whereas the generator given in~\cite{Kimura:2015rgi} is
associated to a Fock module, which is not of finite type. It should be
possible to write down the relations of the $W$-algebra associated to
DIM algebra \emph{explicitly} in the basis corresponding to
\emph{finite} type modules.

Plane partitions naturally arise in the \emph{crystal melting} models
of (refined) topological vertex $C_{\lambda \mu \nu}(\mathfrak{q},
\mathfrak{t})$~\cite{ref, crystal}. Moreover, the matrix
model~\eqref{eq:32} actually reduces to the refined crystal melting
partition function in the NS limit, i.e.\ $q \to 1$ (the remaining
parameters $t_1$, $t_2$ become $\mathfrak{q}$ and
$\mathfrak{t}^{-1}$). The only missing ingredients are the non-empty
asymptotics of the plane partitions featuring in the melting crystal
model. We plan to investigate this connection in the future. We would
also like to point out one mysterious phenomenon along this
direction. Refined topological vertex $C_{\lambda \mu
  \nu}(\mathfrak{q}, \mathfrak{t})$ can be identified with the
\emph{intertwiner} of Fock representations of DIM
algebra~\cite{AFS}. Simultaneously, the Baxter equations obtained from
the corresponding matrix model in the NS limit are relations in the
\emph{representation ring} of DIM algebra. How to make a direct
connection between an intertwiner acting between states of the
representations and a relation between the products of representation
spaces?

Very recently a connection between \emph{solid} (i.e. $4d$) partitions
and gauge theory was put forward in~\cite{Nekrasov:2017cih}. It would
be interesting to understand the relation between our results an those
of~\cite{Nekrasov:2017cih}.

\section*{Acknowledgements}

The author is grateful to M.~Bershtein, B.~Feigin, I.~Frenkel and
S.~Mironov for discussions. The author is supported by the ERC-STG
grant 637844-HBQFTNCER, by the INFN and by the RFBR grants
16-02-01021, 16-51-45029-Ind, 15-51-52031-NSC, 16-51-53034-GFEN,
17-51-50051-YaF.

\end{document}